\documentstyle[twocolumn,twoside,floats,prb,aps,psfig]{revtex}

\begin{document}
\newcommand{\beq}{\begin{equation}}
\newcommand{\eeq}{\end{equation}}
\newcommand{\degree}{$^{\rm\circ} $}
\newcommand{\pcite}{\protect\cite}
\newcommand{\pref}{\protect\ref}

\title{The Physical Origin of Intrinsic Bends in Double Helical DNA}

\author{Alexey K. Mazur}

\address{Laboratoire de Biochimie Th\'eorique, CNRS UPR9080\\
Institut de Biologie Physico-Chimique\\
13, rue Pierre et Marie Curie, Paris,75005, France.\\
FAX:+33[0]1.58.41.50.26. Email: alexey@ibpc.fr}

\date{\today}
\maketitle

\begin{abstract}

The macroscopic curvature induced in the double helical B-DNA by regularly
repeated adenine tracts (A-tracts) is a long known, but still
unexplained phenomenon. This effect plays a key role in DNA studies
because it is unique in the amount and the variety of the available
experimental information and, therefore, is likely to serve as a gate
to the unknown general mechanisms of recognition and regulation of
genome sequences. We report the results of molecular dynamics
simulations of a 25-mer B-DNA fragment with a sequence including three
A-tract phased with the helical screw. It represents the first model
system where properly directed static curvature emerges spontaneously
in conditions excluding any initial bias except the base pair sequence.
The effect has been reproduced in three independent MD trajectories of
10-20 ns with important qualitative details suggesting that the final
bent state is a strong attractor of trajectories form a broad domain
of the conformational space. The ensemble of curved conformations,
however, reveals significant microscopic heterogeneity in contradiction
to all existing theoretical models of bending.  Analysis of these
unexpected observations leads to a new, significantly different
hypothesis of the possible mechanism of intrinsic bends in the double
helical DNA.

\end{abstract}

\section*{Introduction}

The intrinsic sequence dependent curvature of the DNA molecule is
likely to be involved in fundamental mechanisms of genome regulation.
The possibility of strong static bends in the B-DNA double helix has
been proven for sequences containing regular repeats of $\rm A_nT_m,\
with\ n+m>3$, called A-tracts \cite{Crothers:99}.  This effect plays
an important role in DNA studies because it is unique in the amount
and the variety of the available experimental information and,
possibly, it can serve as a gate to the unknown general mechanisms of
recognition and regulation of genome sequences. Every A-tract deviates
the helical axis in a locally fixed direction by approximately
18\degree, and, if the A-tracts are repeated in phase with the helical
screw, a macroscopic curvature emerges. The effect was first
noticed and identified in restriction fragments from the kinetoplast
body of {\em Leishmania tarentolae} \cite{Marini:82,Wu:84}, and
confirmed by electric birefringence decay \cite{Hagerman:84} and
electron microscopy \cite{Griffith:86}. A large variety of interesting
information has been obtained by biochemical methods. It appeared that
the double helix bends towards the minor grooves of A-tracts
\cite{Koo:86,Zinkel:87}. The curvature is reduced with the temperature
above 40\degree\ and in high salt, but for some sequences it is
increased in presence of divalent metal ions \cite{Diekmann:85}. It
depends upon the length and composition of A-tracts as well as on
sequences between them \cite{Koo:86}. Detailed analysis of these
results can be found in comprehensive reviews published in different
years
\cite{Crothers:99,Diekmann:87a,Hagerman:90,Crothers:90,Crothers:92,Olson:96}.

According to many independent experimental observations, the structure
of A-tract sequences should differ significantly from the ``random''
B-DNA. It is well established that, in solution, the poly-dA double
helix is overwound to a twist of around 36\degree\ from around
34\degree\ of a random sequence \cite{Peck:81,Rhodes:81,Strauss:81}.
The models constructed from fiber diffraction data suggest
consistently that the poly-dA double helix is characterized by a very
narrow minor groove and a high propeller twist
\cite{Aymami:89,Lipanov:90,Chandrasekaran:92}.  Yet another
distinction is an apparently large negative inclination of base pairs
\cite{Lipanov:90}. Several A-tracts avaliable in single crystal
structures of B-DNA oligomers have irregular conformations, but
exhibit similar trends toward their centers
\cite{Dickerson:81a,Nelson:87,DiGabriele:89,Edwards:92,DiGabriele:93}.
Even though the curvature is apparently caused by A-tracts, in Xray
structures, A-tracts look generally less prone to bending than other
sequences. Some indirect observations also support this view, notably,
poly-dA fragments move faster than random DNA in gel migration assays
\cite{Koo:86} and avoid wrapping around nucleosome particles
\cite{Simpson:79,Rhodes:79}.

In spite of a large body of the experimental information accumulated
during the last twenty years, the possible physical mechanism of this
effect remains unclear. Since every base pair in a stack interacts
only with the two neighbors, any sequence specificity in the DNA
structure should mainly depend upon the stacking interactions in one
base pair step. Non-local effects are also possible, however, due to
base-backbone interactions and propagation of correlations along the
backbone. The initial experimental data on A-tract bending were
interpreted in terms of two alternative mechanisms, namely, the wedge
model \cite{Trifonov:80} and the junction model \cite{Levene:83}. Both
had to be modified significantly as and when new experimental data
appeared and some other theories were discussed as well. The
possible mechanisms of bending considered in the literature
will be discussed below. Here we note only that none of them explains
all experimental data and can be definitely
preferred \cite{Crothers:99}. The overall pattern has been
additionally complicated when it was found that certain non A-tract
sequences also exhibit distinguishable curvature \cite{Bolshoy:91}.

Here we report new results obtained in free unbiased molecular
dynamics simulations of a DNA oligomer with phased A-tracts. We
managed to find computational conditions in which stable sequence
dependent static curvature emerges spontaneously in good agreement
with experimental observations in terms of both the bending direction
and magnitude. It is found that three independent long time
trajectories converge to conformations with similar bends, but rather
different local structural parameters, in evident contradiction with
the common views of the origin of curvature. Analysis of these
discrepancies leads to a new, significantly different hypothesis of
the possible mechanism of intrinsic bends in the double helical DNA.

\subsection*{Theoretical Background}

As a theoretical problem, the phenomenon of the sequence dependent DNA
bending presents a challenge in many aspects similar to the protein
folding problem. In order to understand the underlying physical
mechanism, one has to analyze terribly noisy experimental data, with
the noise being due to the biological diversity and, therefore,
unremovable. If and when the physical mechanism is understood one will
have to tackle this diversity again, because it will be necessary to
analyze specific DNA sequences. The atom level molecular modeling is
virtually the only theoretical approach that is potentially able to
treat these difficulties. Ideally, we would like to have a model where
the base pair sequence represents the only initial bias towards a
specific conformation. If it could reproducibly yield curved DNA
conformations in agreement with experimental data we should be able
disclose the mechanism of bending in the model, and hope that a similar
mechanism takes place in the nature.

The foregoing scheme, however, is too difficult and, until now, most
of the modeling studies used other strategies. Much has been learned
about the DNA bending mechanics by using energy minimization
\cite{Zhurkin:79,Kitzing:87,Chuprina:88,Sanghani:96} and Monte Carlo
\cite{Zhurkin:91}. Unfortunately, the possibilities of such studies
are limited by the multiple minima problem especially when it is
necessary to take into consideration specific interaction with solvent
molecules. The proposed alternative strategies commonly involved some bias
towards specifically bent conformations introduced either explicitly,
by imposing restraints, or implicitly, by choosing particular initial
conditions, which made impossible unequivocal conclusions concerning
the possible mechanism of bending. In the recent years, owing to the
progress achieved in improving the full atom force fields,
multi-nanosecond free MD simulations of DNA became feasible
\cite{AMBER94:,MacKerell:95}. A few such studies of phased A-tract
sequences have been already reported \cite{Young:98,Sprous:99}. It has
been demonstrated that, without any {\em a priori} bias, the DNA
double helix bends anisotropically, and certain sequence specific
features of A-tracts were at least qualitatively reproduced. At
present, the free MD simulations represent the most promising line of
research in this field, and we continue it here by using the recently
proposed minimal model of B-DNA \cite{Mzjacs:98,Mzlanl:99}.

The minimal B-DNA consists of a double helix with the minor groove
filled with explicit water. Unlike the more widely used models, it
does not involve explicit counterions and damps long range
electrostatic interactions in a semi-empirical way by using distance
scaling of the electrostatic constant and reduction of phosphate
charges. We have earlier found that the minimal model gives B-DNA
conformations which better compare with experimental structures than
DNA structures obtained with other computational methods. Notably, it
is free from a systematic negative bias of the average twist observed
in simulations with full hydration and explicit counterions
\cite{Mzjacs:98}. This factor is likely to be involved in DNA bending
because, as noted above, A-tracts are overwound with respect to the
average B-DNA. For the standard test case of the EcoRI dodecamer
structure, the dynamics of the minimal model reproducibly converged to
structures slightly bent towards the minor groove which was narrowed
in excellent agreement with the single crystal conformation
\cite{Mzjacs:98,Mzlanl:99}. All these preliminary observations
suggested that the minimal model was a good choice for studying the DNA
bending induced by A-tracts.

We report here simulations of the bending dynamics of a 25-mer B-DNA
fragment. Its sequence, $\rm AAAATAGGCTATTTTAGGCTATTTT$, has been
constructed after many preliminary tests with shorter sequence
motives, and it includes three A-tracts separated by one helical turn.
Our general strategy came out from the following considerations.
Although the A-tract sequences that induce the strongest bends are
known from experiments, probably not all of them would work in
simulations. There are natural limitations, such as the precision of
the model, and, in addition, the limited duration of trajectories may
be insufficient for some A-tracts to adopt their specific
conformation. Also, there is little experimental evidence of static
curvature in short DNA fragments, notably, one may well expect the
specific A-tract structure to be unstable near the ends. That is why
we did not simply take the strongest experimental ``benders'', but
looked for sequence motives that in calculations readily adopt the
characteristic local structure, with a narrow minor groove profile and
high propeller twist, both in the middle and near the ends of the
duplex. The complementary duplex $\rm AAAATAGGCTATTTTAGGCTATTTT$ has
been constructed by repeating and inverting one such motive.

Our goal was to find sequences that would appear statically bent in
these conditions. It means that, in dynamics, the structure should
fluctuate around a state with a distinguishable bend and a definite
bending direction. Since any MD simulation is limited in time, there
is no way to prove rigorously that some specific conformation is
representative. Some degree of confidence can be achieved, however, if
independent trajectories are able converge to the same state, and this
is exactly what we tried to obtain. We found important to place
A-tracts at both ends because of the two reasons. First, we could
study only short DNA fragments, therefore, it was preferable to place
A-tracts at both ends in order to maximize the possible bend. Second,
in calculations with short sequences, it is important to be sure that
the boundary conditions are correct. Since the A-tracts have a
characteristic local structure, the boundary conditions could be at
least qualitatively verified, which would not be the case for GC-rich
sequences.

\section*{Results}

Three long MD trajectories were computed for a complementary DNA duplex
with the sequence $\rm AAAATAGGCTATTTTAGGCTATTTT$.
The model system employed was same in
all three simulations, with only the starting states varied. The first
trajectory referred to below as TJBa started from the fiber canonical
B-DNA structure \cite{Arnott:72} and continued to 10 ns.  When it was
found that TJBa converged to a statically bent conformation, in good
qualitative agreement with expectations based upon experimental data,
another trajectory (TJBb) was computed in order to verify the
reproducibility of the results. It started with random velocities from
a re-minimized straight conformation taken from the initial phase of
TJBa and was also continued to 10 ns. Simultaneously, in order to
remove any initial bias implicitly involved in the choice of the
starting state, the third trajectory (TJA) was obtained which started
form the fiber canonical A-DNA conformation \cite{Arnott:72}.
Initially, we computed 10 ns of TJA and found that it sampled
conformations rather dissimilar from those observed in TJBa and TJBb.
A careful analysis revealed, however, certain slow structural trends
and prompted us to continue TJA to 20 ns. The first two trajectories
(TJBa and TJBb) have been the subject of our initial report \cite{Mzlanl:00}.
Therefore, below we describe in detail only TJA and use the
corresponding data from TJBa and TJBb in comparisons. The structures
referred to as the final MD states are the conformations averaged over
the last nanosecond of the corresponding trajectory. The detailed
computational protocols are described in Methods.

\subsection*{All Three Trajectories Converge to Similar Structures
within B-DNA Family}

\begin{table}[t]\caption{\label{Trmsd} Nonhydrogen Atom
rmsd (\AA) between standard and computed structures. The upper
and the lower triangles show results for all and for the middle
11 base pairs, respectively}
\begin{tabular}[t]{|dddddd|}
 & A-DNA\tablenotemark[1]
    & B-DNA\tablenotemark[1]
       & TJA\tablenotemark[2]
          & TJBa\tablenotemark[2]
              & TJBb\tablenotemark[2]\\
\hline
A-DNA\tablenotemark[1]  & 0.0 &  8.62 &  9.18 & 8.62 & 7.98\\
B-DNA\tablenotemark[1]  & 5.84 &  0.0 &  3.13 & 2.85 & 3.28\\
TJA\tablenotemark[2]   & 6.15 &  1.96 &  0.0 & 2.90 & 2.92\\
TJBa\tablenotemark[2]  & 5.41 &  1.35 &  1.44 & 0.0 & 1.79\\
TJBb\tablenotemark[2]  & 5.66 &  1.83 &  1.39 & 1.32 & 0.0\\
\end{tabular}

\tablenotetext[1]{
Fiber canonical DNA conformations constructed
from the published atom coordinates \cite{Arnott:72}.
}

\tablenotetext[2]{
MD conformations averaged over the last nanosecond of the
corresponding trajectory.
}

\end{table}

Table \ref{Trmsd} presents atom rmsd comparison between the final MD
states of the tree trajectories and canonical A and B-forms of this
25-mer duplex. All computed structures are clearly different from the
standard A-DNA, even though for TJA it was the starting point.
Moreover, the TJA final state appears to be the less similar of the
three, demonstrating that the trajectories were not trapped
kinetically in the vicinities of their starting points. At the same
time, the final MD states are evidently close to the canonical B-DNA.
When only the central undecamer is considered, the rmsd values are in
the same range as those observed in our earlier MD simulations of
dodecamer duplexes \cite{Mzjacs:98,Mzlanl:99}.
They are low, and the three computed
conformations seem to form a single cluster around the canonical
B-DNA. The rmsd naturally increases with the helix length, but it
should be noted that the somewhat larger values obtained for the whole
structures are much lower than ever observed in free MD simulations of
long DNA helices \cite{Young:98}. It appears, however, that, if taken
as a whole, the TJA state is as close to the canonical B-DNA as to the
TJBa and TJBb states, while the latter two are yet closer to each
other.

\begin{figure}
\centerline{\psfig{figure=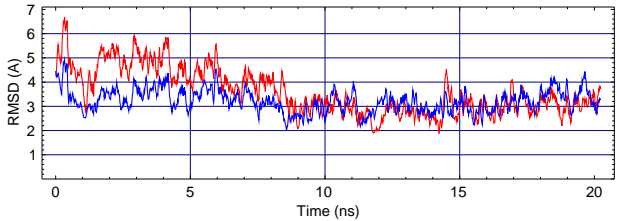,height=3cm,angle=0.}}
\caption{\label{Ftcrmsd}
Kinetics of the structural convergence for TJA. The time
dependence of the nonhydrogen atom rmsd from the canonical B-DNA
conformation (blue) and the final TJBa state (red)
are shown. Both traces were smoothed by averaging with a window
of 75 ps.
}
\end{figure}

\begin{table*}[t]\caption{\label{Thlpa} Some Structural Parameters
of Standard and Computed DNA Conformations}
\begin{tabular}[t]{|ccccccc|}
           & Xdisp\tablenotemark[1]
               & Inclin\tablenotemark[1]
                    & Rise\tablenotemark[1]
                       & Twist\tablenotemark[1]
                           & Bend\tablenotemark[2] & Bending\tablenotemark[3]\\
           &    &    &   &    & angle & direction\\
\hline
A-DNA\tablenotemark[4] & -5.4 & +19.1 & 2.6 & 32.7 & 0.0 & -- \\
B-DNA\tablenotemark[4] & -0.7 & -6.0 & 3.4 & 36.0 & 0.0  & -- \\

TJA\tablenotemark[5] & -0.7|-0.3 & +0.0|-1.2 & 3.5|3.5 & 34.3|34.8 & 35.8|30.4 & -136.8|+110.4 \\
TJBa\tablenotemark[5] & -1.0|-0.3 & -6.9|-4.5 & 3.6|3.5 & 34.1|34.8 & 13.2|29.5 & 22.3|42.9   \\
TJBb\tablenotemark[5] & -1.0|-0.7 & -4.6|-1.8 & 3.5|3.5 & 34.3|34.2 & 18.4|28.4 & 35.6|53.7   \\
\end{tabular}

\tablenotetext[1]{
Sequence averaged values computed with program Curves \cite{Curves:}.
}

\tablenotetext[2]{
The angle between the two terminal vectors of the optimal helical axis.
}

\tablenotetext[3]{
The bending direction is characterized by the
angle between the plane of the optimal helical axis and the $xz$ plane
of the local DNA coordinate frame constructed in the center of the
duplex. According to the Cambridge convention \cite{Dickerson:89} the
local $x$ direction points to the major DNA groove along the short axis
of the base-pair, while the local $z$ axis direction is adjacent to
the optimal helicoidal axis. Thus, a zero angle between
the two planes corresponds to the overall bend to the
minor groove exactly at the central base pair. 
}

\tablenotetext[4]{
Fiber canonical DNA conformations constructed
from the published atom coordinates \cite{Arnott:72}.
}

\tablenotetext[5]{
MD conformations averaged over one nanosecond intervals.
The two numbers correspond to the first and the last nanosecond,
respectively.
}

\end{table*}
The kinetics of the structural convergence in terms of atom rmsd is
illustrated in Fig. \ref{Ftcrmsd} for TJA. It is seen that
the trajectory rapidly went from the initial A-DNA conformation towards the
B-DNA form and, after the equilibration, the rmsd from B-DNA have
already lost a half of the initial 8.6 \AA. Starting from the second
nanosecond it fluctuated between 2 and 4 \AA. The corresponding
kinetics for TJBa and TJBb were very similar except for the initial
fall of the rmsd value \cite{Mzlanl:00}. The rmsd from the TJBa state
also shown in Fig. \ref{Ftcrmsd} falls down
 to similar final values, but exhibits a
somewhat different kinetics. Namely, an overall negative drift occurred
during the first ten nanoseconds followed by random fluctuations
during the second half of the trajectory. One may say, therefore, that
TJA first quickly traveled from A-DNA towards the B-DNA family and
next slowly refined its position within this family coming closer to
other computed structures. This refinement was not complete, however,
since, according to Table \ref{Trmsd}, the final TJA-TJB difference is
still larger than that between TJBa and TJBb. Figure \ref{Ftcrmsd}
suggests that a more accurate convergence, if possible, would require
much longer time.

Table \ref{Thlpa} compares a few representative structural parameters
of MD conformations with the corresponding standards. Already after
the first nanosecond even TJA gave the helicoidals corresponding to
the B-DNA family, and they exhibited no systematic change afterwards.
All three final MD states have an overall bend of around 30\degree .
The bending direction is somewhat different between TJA and TJB, which
is the main cause of the corresponding residual difference in terms of
atom rmsd. Table \ref{Thlpa} indicates that in all three trajectories
both the magnitude and the direction of the bends changed
significantly, and that very large variations in the bending direction
apparently occurred in TJA. Thus, the slow rmsd kinetics considered
above appear to be largely due to the bending of the double helices
whereas the contribution from the variations of the helical parameters
looks minor.

\begin{figure}
\centerline{\psfig{figure=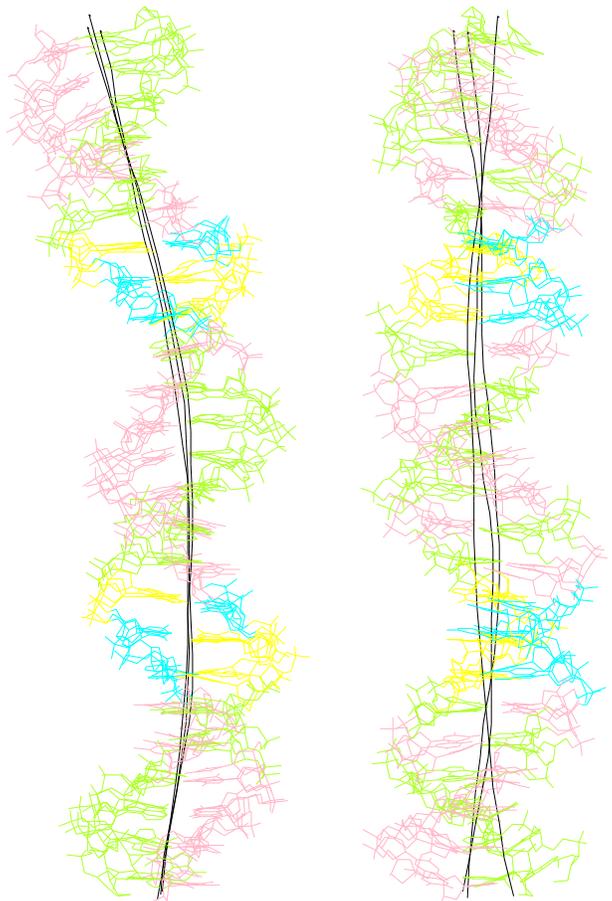,height=12cm,angle=0.}}
\caption{\label{F3str}
The final MD states of the three trajectories
superimposed with optimal helical axes shown by curved lines. Two
perpendicular views are shown. The superimposed structures were
rotated to minimize the divergence of the projections of
the helical axes in the left hand view. Different nucleotides are
coded by colors, namely, A - green, T - red, G - yellow, C - blue.
}
\end{figure}

Figure \ref{F3str} shows the three last nanosecond average structures
superimposed. They all are evidently curved, with the bends being
nearly planar in each structure. In agreement with Table \ref{Thlpa},
the TJA bending plane slightly deviates from the other two. The bending planes
intersect the minor groove in five points which alternate between the
inside and the outside edges of the bend, and in each case the the
three A-tracts appear at the inside edge. The tracts are approximately
phased with the helical turn, but, since the lower one is inverted
with respect to the other two, its 5' end is phased with the 3' ends
of the other. The three inside intersection points are shifted within
the A-tracts from their middle towards the 3' end of the upper two and
the 5' end of the lower one. On the other hand, the minor grooves of
the two AGGC tetraplets appear at the outside edge of the curved axis,
and it is readily seen that the minor groove is widened here,
especially at the upper tetraplet.

\subsection*{Quasi-Regular Rotation of The Bending Plane in TJA}

\begin{figure}
\centerline{\psfig{figure=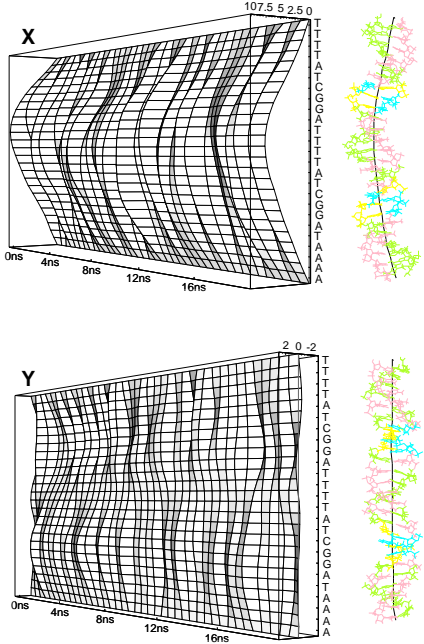,height=9cm,angle=0.}}
\caption{\label{Faxsrf}
The time evolution of the overall shape of the optimal helical axis in
TJA. A best fit axis of coaxial cylindrical surfaces passing through
sugar atoms was computed for conformations stored with a 2.5 ps
interval. In all cases presented here this axis was close to that
produced by the Curves algorithm \pcite{Curves:}. The curved DNA axis
is rotated with the two ends fixed at the OZ axis to put the middle
point at the OX axis. Note that this procedure does not keep the
structures superimposed, that is the same curved axis can correspond
to different bending directions. The axis is next characterized by two
perpendicular projections labeled X and Y. Any time section of the
surfaces shown in the figure gives the axis projection averaged over a
time window of 150 ps. The horizontal deviation is given in
angstr\"oms and, for clarity, its relative scale is two times
increased with respect to the true DNA length. Shown on the right are
the two perpendicular views of the last one-nanosecond-average
conformation in the orientation corresponding to that in the surface
plots at the end of the trajectory.
}
\end{figure}

The two surface plots in Fig. \ref{Faxsrf} exhibit the time evolution
of the shape of the helical axis for TJA. It is seen that the molecule
was strongly bent after the initial equilibration, which was not
observed in case of TJBa and TJBb \cite{Mzlanl:00}. One should note that
considerable initial deformation of the double helix is common for
trajectories starting from the A-DNA conformation.
Apparently, the molecule is stressed because the transition to the
B-form occurs in these conditions during unphysically short time with
much energy released. During the next few nanoseconds the bend reduced
and the axis acquired a more complex shape with wound profiles in both
projections. After the fifth nanosecond the bending became more planar, with
much smaller curvature in Y projection. A planar bent may just mean that
the helical axis is kinked in a single point or, alternatively, a
lager number of local bends are properly directed. Figure \ref{Faxsrf}
indicates that there is probably a mixture of these two effects.
During the last few nanoseconds the axis had one stable bending point
shifted upwards from the middle while another bend in the lower half
emerged from time to time. The two bends were slightly misaligned,
therefore, the overall bending plane rotated a little when the second
bend emerged, and, in the Y projection, one sees alternation of
straight and S-shaped profiles.

\begin{figure}
\centerline{\psfig{figure=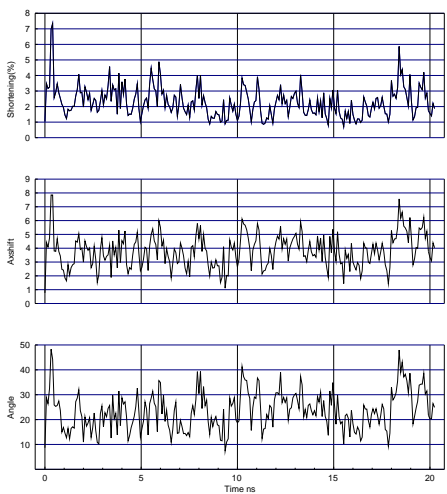,height=9cm,angle=0.}}
\caption{\label{Ftcbnd}
The time evolution of the magnitude of bending in TJA. The bending
angle is measured between the two ends of the helical axis. The shift
parameter is the average deviation of the helical axis from the
straight line between its ends. The shortening is measured as the
ratio of the lengths of the curved axis to its end-to-end distance
minus one.
}
\end{figure}

Figure \ref{Ftcbnd} displays kinetics of several quantitative measures
of the magnitude of bending. The three parameters used, namely, the
total angle, the shortening, and the average shift of the curved axis,
all exhibit a coherent pattern of fluctuations, which locally correlates
also with the rmsd from the canonical B-DNA (see Fig. \ref{Ftcrmsd}).
This indicates that
they all are produced by the same motion, namely, the axis bending.
Comparison of the data in Figs. \ref{Faxsrf} and \ref{Ftcbnd} with
similar plots earlier reported for TJBa and TJBb \cite{Mzlanl:00} reveals
little difference except the already mentioned initial deformation and
the absence of a stable bend between the two lower A-tracts.
Accordingly, TJBa and TJBb showed a somewhat stronger bending, with
one-nanosecond average values usually beyond 35\degree . In TJA, after
the initial strong temporary bending, a comparable magnitude has been
reached only during the last four nanoseconds.

\begin{figure}
\centerline{\psfig{figure= 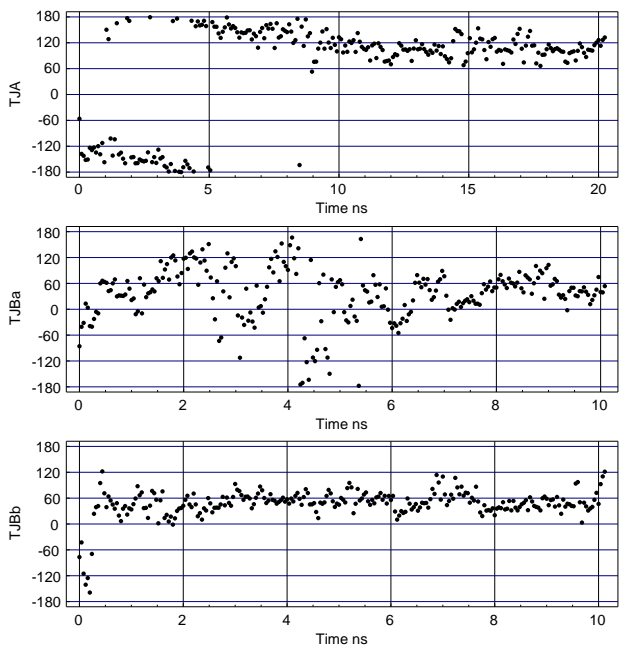,height=9cm,angle=0.}}
\caption{\label{Ftcdir}
The time evolution of the bending direction in the tree trajectories.
The bending direction is characterized
by the angle (given in degrees) between the
X-projection plane in Fig. \pref{Faxsrf} and the $xz$ plane of the
local DNA coordinate frame at the center of the duplex. It is
constructed according to the Cambridge convention
\pcite{Dickerson:89}, namely, the local $x$ direction points to the
major DNA groove along the short axis of the base-pair, while the
local $z$ axis direction is adjacent to the optimal helicoidal axis.
Thus, a zero angle between the two planes corresponds to the overall
bend to the minor groove exactly at the central base pair.
}
\end{figure}

There is, however, a striking difference between TJA and the other two
trajectories in the dynamics of the bend direction which is exhibited
in Fig. \ref{Ftcdir}. Both in TJBa and TJBb the final bending
direction occurred early in the trajectories and remained quite stable
although the molecule sometimes straightened producing broad
scattering of points in Fig. \ref{Ftcdir}. In contrast, during
the first ten nanoseconds of TJA, the bending plane made almost a half turn
with respect to the coordinate system bound to the molecule. It means
that a transition occurred between the oppositely bent conformations,
but, as seen in Fig. \ref{Faxsrf}, the straight one was avoided.
This rotation was very steady, almost regular.
It gradually slowed down becoming indistinguishable in the
last five nanoseconds. After this transition the directions of
the bends in the three trajectories became much closer,
and this quasi-regular motion is apparently responsible for the
slow drift of the rmsd from the TJBa state in Fig.
\ref{Ftcrmsd}. The overall amplitude of this motion was
around 150\degree , that is the initial strong bend noticed in
Fig. \ref{Faxsrf} was nearly opposite to that finally established.

\subsection*{The Rotation of the Bending Plane is Not Energy Driven}

\begin{figure}
\centerline{\psfig{figure=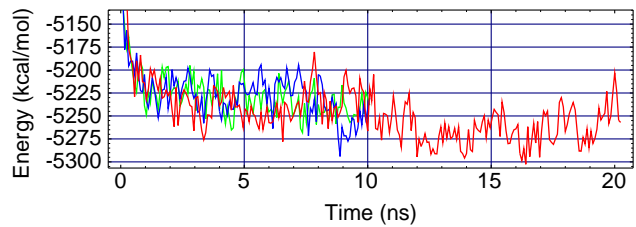,height=3.5cm,angle=0.}}
\caption{\label{Ftcey}
The time dependence of the potential energy in three
different trajectories. The color coding is: TJA -- red, TJBa -- green
and TJBb -- blue.
}
\end{figure}

The overall character of motion revealed in Fig. \ref{Ftcdir},
namely, the steady rotation of the bending plane, looks
strange and counter-intuitive. {\em A priori}, we would rather expect
to obtain random sampling of different bending directions, with the correct
one statistically preferred due to its lower energy. The apparent
quasi-regular dynamics exhibited in Fig.
\ref{Ftcbnd} might mean that our trajectory represents a downhill
motion along a valley on a potential energy surface. Its steep borders
would separate bent conformations from the straight one, with the
bottom of this valley slightly inclined towards the preferred bending
direction. In this case all bent conformations, including incorrect
bends, should have been lower in energy than the straight one.

Figure \ref{Ftcey} displays the time evolution of the potential energy
in all three trajectories. It is seen that the energy dropped during
the first nanosecond and later remained stable. No clear correlation
is seen between the instantaneous magnitude of bending displayed in
Fig. \ref{Ftcbnd} and the potential energy, therefore, one cannot say
that straight states have significantly different energies than the
bent ones. Neither can we claim that the preferred bending direction
is characterized by lower energy values than other bends. In Fig.
\ref{Ftcey}, a slight decrease in energy is observed during the second
half of TJA, but it occurred when the regular rotation of the bending
plane has essentially finished. On the other hand, the lowest energy
during the first half of the trajectory was observed at around 3.2 ns
when the bending direction was completely different. Note also that,
during the first ten nanoseconds, the traces of TJBa and TJBb go above
the last one, although in these cases the correct bending direction
has already established. We have to conclude, therefore, that the
simple energetic al explanation of the observed effect does not work.

\subsection*{The Minor Groove Profiles of Converged Structures
Are Similar But Not Identical}

\begin{figure}
\centerline{\psfig{figure=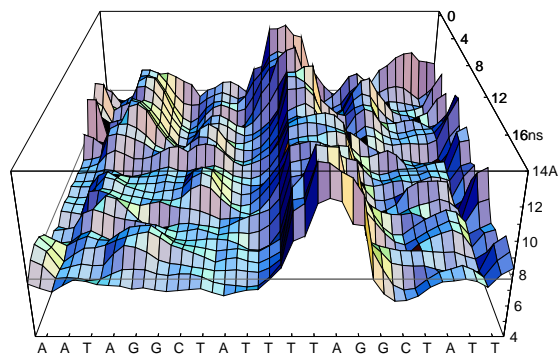,height=5cm,angle=0.}}
\caption{\label{Fmgkt}
The time evolution of the profile of the minor groove in TJA. The surface is
formed by 150 ps time-averaged successive minor groove profiles, with
that on the front face corresponding to the final DNA conformation.
The groove width is evaluated by using space traces of C5' atoms as
described elsewhere \pcite{Mzjmb:99}. Its value is given in angstr\"oms
and the corresponding canonical B-DNA level of 7.7 \AA\ is marked by
the straight dotted lines on the faces of the box.
}
\end{figure}

The surface plot in Fig. \ref{Fmgkt} exhibits the evolution of the
profile of the minor groove during TJA. The initial A-DNA conformation
is characterized by a uniformly wide minor groove of 13.6 \AA. It is
seen that after the equilibration period the groove was much narrower,
but still wider than in the canonical B-DNA model. Moreover, a complex
profile have emerged with three local widenings at A-tracts, which is
exactly opposite to the expectations. The two terminal widenings
reduced during the first ten nanoseconds whereas the maximum of the
middle one gradually shifted from its 3' end to 5' end. This shift
evidently accompanies the rotation of the bending plane described
above.

One can note that the maximal widening of the minor groove moved for
only 2-3 base pair steps, which is less than a 150\degree rotation
seen in Fig. \ref{Ftcdir}. It appears that, in fact, the maximal
widenings and narrowings in the minor groove profile do not always
correspond to the direction of local bends. The initial bend
was directed towards the minor groove of the upper TAGG
tetraplet where the minor groove was narrowed. The two neighboring
widenings are shifted by three base pairs only and they appear at the
opposite sides of the bending plane which is approximately collinear to
the pseudodiad axis at the center of the middle ATT triplet.
In contrast, in the last structure shown
in Figs. \ref{F3str} and \ref{Faxsrf} the bending
plane passes exactly through the maximum widening of the minor groove.
The overall rotation, therefore, corresponds to approximately four
base pair steps which gives the observed turn by 150\degree . It can
be noted, finally, that although Fig. \ref{Ftcdir} indicates that the
bending stabilized after ten nanoseconds, the profile of
the minor groove in  Fig. \ref{Fmgkt} continues
to evolve slowly till the very end of the trajectory.

\begin{figure}
\centerline{\psfig{figure=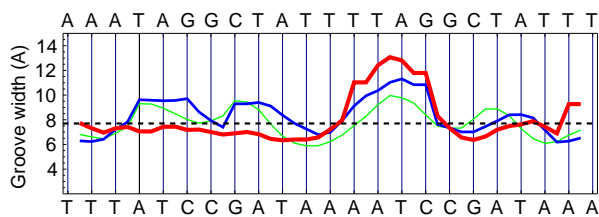,height=3cm,angle=0.}}
\caption{\label{Fmgcmp}
The profiles of the minor groove in the last nanosecond
average conformations from TJA, TJBa and TJBb. The groove width
is evaluated by using the C5' traces as described elsewhere
\pcite{Mzjmb:99}. The dotted line marks the groove width of the
canonical B-DNA structure \pcite{Arnott:72}. The color coding is:
TJA -- red, TJBa -- green and TJBb -- blue.
}
\end{figure}

Figure \ref{Fmgcmp} displays the minor groove profiles of the last
average structures from the three trajectories. For TJBa and TJBb
their kinetics was detailed in our first report \cite{Mzlanl:00}, and we only
note here that the corresponding profiles shown in Fig. \ref{Fmgcmp}
established during the first two nanoseconds and showed little
variations afterwards \cite{Mzlanl:00}. The three traces evidently exhibit a
certain similarity, but do not coincide. The TJBa and TJBb grooves
have the same number of local narrowings and widenings which differ
slightly between the two both in amplitude and in position. The TJA
profile is similar in the right-hand half of the figure. One can
notice that the change from TJBa to TJBb and next to TJA involves the
growing widening at the TTAG tetraplet accompanied by a shift of the
secondary maximum, and looks rather regular and concerted. At the
opposite half of the structure, the TJA conformation shows a narrow
minor groove without significant modulations of the width. This
difference may be related with the smaller magnitude of the bending in
the case of TJA where the second bending point appeared from time to
time only and was less significant than in the other two trajectories.

\subsection*{Key Helicoidal Parameters Exhibit Consistent
Regular Patterns Only after Window Averaging}

\begin{figure}
\centerline{\psfig{figure=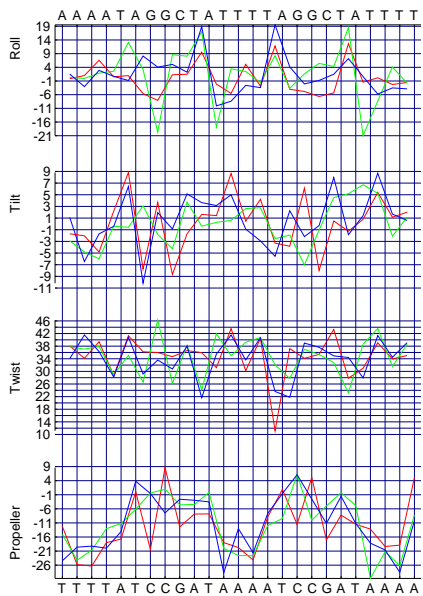,height=8cm,angle=0.,%
bbllx=0pt,bblly=-14pt,bburx=253pt,bbury=300pt,clip=t}}
\caption{\label{Fhlpa}
Sequence variations of some helicoidal parameters in the final states
of TJA, TJBa and TJBb. The sequence of the first strand is shown on
the top in 5' -- 3' direction. The complementary sequence of the
second strand is written on the bottom in the opposite direction. All
parameters were evaluated with the Curves program \pcite{Curves:} and
are given in degrees. The color coding is: TJA -- red, TJBa -- green
and TJBb -- blue.
}
\end{figure}

Figure \ref{Fhlpa} shows variation of some helicoidal parameters along
the duplex in the three structures. The two inter base pair
parameters, namely, roll and tilt, are most often quoted in the
literature in relation to the static DNA curvature. If one first takes
an ideal straight column of stacked parallel base pairs and next
introduce a non-zero roll value at a certain step, the structure will
bend at this step towards the major groove, if the roll is positive,
and to the minor groove if it is negative. A similar experiment with
the tilt value would result in bending in the perpendicular direction.
It seems obvious that, whatever the physical origin of the curvature,
in a bent double helical DNA, the roll and tilt values must exhibit
systematic variations phased with the helical turn. Moreover, it is
often assumed that for some short DNA sequences certain non-zero roll
and tilt values are strongly preferred energetically, which produces
static bending when they are repeated appropriately.

\begin{figure}
\centerline{\psfig{figure=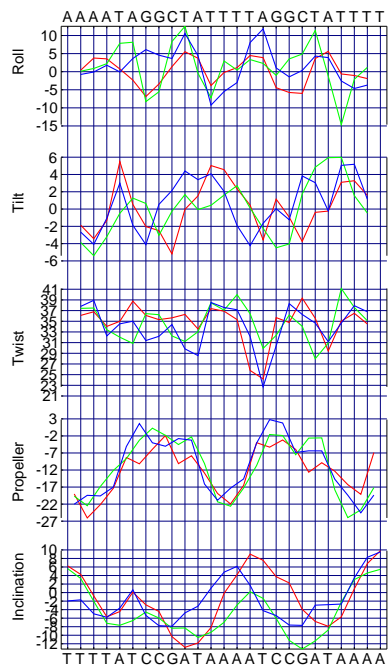,height=9cm,angle=0.,%
bbllx=0pt,bblly=-14pt,bburx=253pt,bbury=300pt,clip=t}}
\caption{\label{Fhlpawn}
Sequence variations of some helicoidal parameters in the final states
of TJA, TJBa and TJBb, with window smoothing applied to the tilt, roll,
twist, and propeller parameters. A sliding window of two base pair steps was
applied to the data in Fig. \pref{Fhlpa}, with the resulting average
value assigned to the middle point. Notation as in Fig. \pref{Fhlpa}.
}
\end{figure}

However surprising, although all three average structures are smoothly
curved, only a few supporting signs for the foregoing paradigm are
readily seen in Fig. \ref{Fhlpa}. For the tilt, the three traces are
very dissimilar and the only feature that repeats is the alteration of
its values between consecutive steps. Namely, if the tilt is low at a
given step it normally goes up at the next one, and {\em vice versa}.
In the three average structures, however, these alterations are
sometimes oppositely phased even in TJBa and TJBb where the
overall structures look particularly similar.

The same is true for the roll and twist although, in these cases, some
clear sequence preferences do exist. Note, for instance, that, in all
four TpA steps, the roll is almost always positive and larger than in the
neighboring steps. Paradoxically, two of these TpA steps occur almost
exactly at the inside edge of the curved axis, that is a high positive
roll accompanies the bending in an exactly opposite direction. This
paradox is readily resolved when one looks at the roll values at the
neighboring steps. A TpA step with a high positive roll is normally
preceded or followed by a step with a low negative roll. The higher is
the maximum, the lower is the neighboring minimum, so that the two
nearly cancel each other. The other two TpA steps are found at the
outside edge of the helical axis and their high roll probably
contributes to bending. However, while a more or less repetitive
pattern is observed around the third TpA step, the first one exhibits
rather dissimilar pictures even for TJBa and TJBb structures which
both have a widened minor groove here. Also, the roll values at the
third TpA step differ considerably between the structures, but do not
correlate with the bending magnitudes.

The twist, tilt, and roll values used for the plots in Fig. \ref{Fhlpa} are
the so called ``global'' parameters from the outputs of the Curves
program \cite{Curves:}. One may argue that they are not
appropriate in the present context since they are computed by using
local directions of already curved optimal helical axis. However, when
``local'' values are used instead, the amplitudes of the alternations
in these profiles are only increased.

The last plot in Fig. \ref{Fhlpa} exhibits the variations of the
propeller twist. Again one sees that its value alternates between
consecutive base pairs, with little phase similarity between the three
structures. At the same time, in this case, a consistent repetitive
pattern is evident, with strong negative propeller values in all
A-tracts. These regular patterns look even more similar than the
structures themselves. For instance, there is no evident difference
between the three traces that would correspond to that in the minor
grove profiles in Fig. \ref{Fmgcmp}.

The apparent jumping alterations of the helicoidal parameters along
the double helix naturally suggests that one should try to smooth them
out by averaging the traces in Fig. \ref{Fhlpa} with a sliding window.
Figure \ref{Fhlpawn} shows the results of such treatment and also
includes the corresponding data for the inclination which was,
however, used without the smoothing. The difference between Figs.
\ref{Fhlpa} and \ref{Fhlpawn} is rather significant. Now all four
helicoidal parameters considered in Fig. \ref{Fhlpa} exhibit regular,
sometimes almost sinusoidal, oscillations. The phasing of these
oscillations with the helical turn, however, is not always evident.
The propeller and the inclination both exhibit approximately 2.5
periods, that is the dominating Fourier component has a wave length of
approximately ten base pairs corresponding to one helical turn.  For
the roll, the dominating wave length apparently corresponds to 5-6
base pairs, that is a half of a helical turn. When different
structures are compared, however, it is seen that only for propeller
the maxima and minima coincide well. A more complex pattern is
observed for the twist, and one can notice a correlation between
its traces and the minor groove profiles shown in Fig. \ref{Fmgcmp}.
Namely, the twist is lower in the widenings of the groove and higher
in its narrowings.

These results suggest that the relationship between the helicoidal
parameters and the bending is complex and cannot be reduced to simple
models of roll-like or tilt-like bends outlined above. Accumulation of
the regular variations revealed in Fig. \ref{Fhlpawn} probably gives
the correct overall bend angles and directions, but neither can be
easily predicted just by looking at these traces.

\subsection*{The Distributions of $\rm\bf B_I\ and\ B_{II}$ Backbone
Conformers in Bent Structures are Surprisingly Dissimilar}

\begin{figure}
\centerline{\psfig{figure=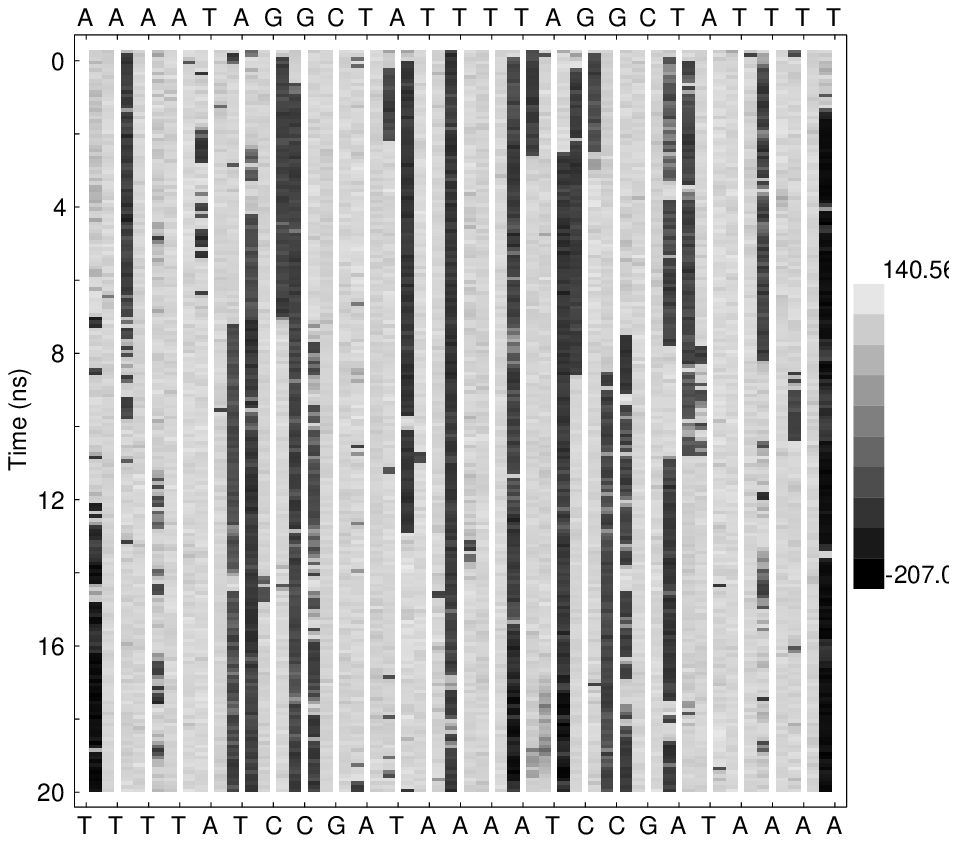,height=8cm,angle=0.}}
\caption{(a)}
\end{figure}\addtocounter{figure}{-1}

\begin{figure}
\centerline{\psfig{figure=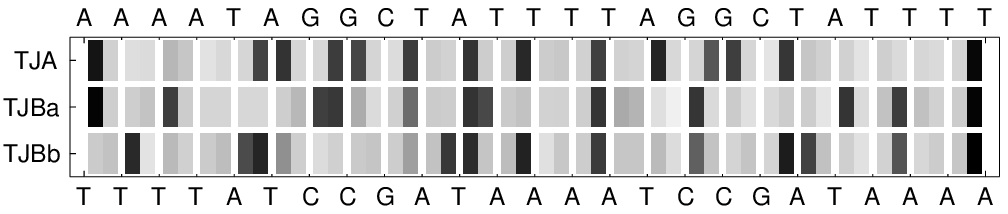,height=2cm,angle=0.}}
\caption{(b)}
\end{figure}\addtocounter{figure}{-1}

\begin{figure}
\caption{\label{FB1.B2}
Dynamics and final distributions of $\rm B_I$ and $\rm B_{II}$
backbone conformers. The B$_{\rm I}$ and B$_{\rm II}$ conformations
are distinguished by the values of two consecutive backbone torsions,
$\varepsilon$ and $\zeta$. In a transition they change concertedly
from (t,g$^-$) to (g$^-$,t). The difference $\zeta -\varepsilon$ is,
therefore, positive in B$_{\rm I}$ state and negative in B$_{\rm II}$,
and it is used in as a monitoring indicator, with the corresponding
gray scale levels shown on the right in plate (a). Each base pair step
is characterized by a column consisting of two sub-columns, with the
left sub-columns referring to the sequence written at the top in 5'-3'
direction from left to right. The right sub-columns refer to the
complementary sequence shown at the bottom. (a) Dynamics of $\rm
B_I\leftrightarrow B_{II}$ backbone transitions in TJA. (b) The
distributions of $\rm B_I$ and $\rm B_{II}$ conformations in the final
states of the three trajectories
}
\end{figure}

In all three trajectories, dynamics of $\rm B_I\leftrightarrow B_{II}$
backbone transitions was qualitatively similar in a few aspects.
Consider Fig. \ref{FB1.B2}a, for instance, where the results are shown
for TJA. The overall pattern reveals rather slow dynamics, suggesting
that MD trajectories in the 10 ns time scale are not long enough to
sample all relevant conformations. A somewhat higher $\rm
B_I\leftrightarrow B_{II}$ activity was observed during the first half
of the trajectory, when the rotation of the bending plane occurred. It is
seen that, in A-tracts, the B$_{\rm II}$ conformers are preferably
found in ApA steps and that they tend to alternate with B$_{\rm I}$
within the same strand. There are many examples of concerted $\rm
B_I\leftrightarrow B_{II}$ transitions, when a given step switches
from $\rm B_{II}\ to\ B_I$ simultaneously with an opposite transition
in one of the neighboring steps. Sometimes three consecutive steps are
involved and, less often, the opposite strand as well. Many $\rm
B_I\leftrightarrow B_{II}$ transitions are reversed within hundreds of
picoseconds, but there are also very long-living conformers and sites
where either $\rm B_I\ or\ B_{II}$ states are preferred. A strong
preference of $\rm B_I$ state is observed for all TpT steps, for
example. However, it seems to be the only case when the effect repeats
at a base pair step level. In a few steps where the $\rm B_{II}$
conformation is preferred this is apparently determined by a broader
sequence context.

The corresponding data for TJBa and TJBb were included in our first
report and they revealed the same qualitative features \cite{Mzlanl:00}. It
was very surprising for us, however, that, in spite of the good
convergence in terms of the overall bent shape of the molecule, the
three trajectories gave rather dissimilar distributions of $\rm B_I\
and\ B_{II}$ conformers along the sequence. Fig. \ref{FB1.B2}b
compares these distribution in the final backbone conformations in the
three trajectories. There are 14 non TpT steps where the $\rm B_I$
conformation is found in all three structures. However, since our
trajectories started from $\rm B_I$ states, this number hardly tells
us something. On the other hand, the number of $\rm B_{II}$ conformers
found in each structure and in each strand is similar and roughly
corresponds to 25\% of phosphate groups. Assuming that the $\rm
B_{II}$ states are evenly distributed in the sequence one gets the
expectation value of 0.75 for the number of cases when the $\rm
B_{II}$ conformer should be found in the same base pair step in all
three structures. The observed number of such sites is three. Note,
however, that they all are found in A-strands of A-tracts where, as
noted above, the $\rm B_{II}$ conformers tend to alternate. This,
together with the strong preference of TpT steps for $\rm B_{I}$,
increases the probability of matching.

These results suggest that the relationship between the bending of
the DNA double helix and the $\rm B_I\leftrightarrow B_{II}$ backbone
transitions, if any, is loose in the sense that a given bent shape
does not impose a fixed $\rm B_{II}$ distribution upon the backbone.

\section*{Discussion}

This study gives the first example of a successful implementation of
the general strategy outlined in Theoretical Background. Namely, we
showed that the minimal model of B-DNA, which is biased only by the
nucleotide sequence, in dynamics, reproducibly converges to a single
state characterized by an ensemble of similar statically bent
conformations. The effect has been demonstrated here for one sequence
only. Moreover, this sequence was specifically constructed rather than
taken from experimental studies.  Nevertheless, the sequence motive
A$_n$TAG used in construction was found in the center of the first
bent DNA fragment studied experimentally \cite{Wu:84}.  In addition,
the character of bending in the computed conformations, notably, its
direction with respect to the A-tracts, and modulations of the groove
width, qualitatively agree with the rules derived from experiments.
These observations validate an attempt to make the next step of the
above strategy, namely, below we try to disclose the mechanism
responsible for the bending within the framework of the minimal model.
We believe that, in spite of the obvious limitations of this model,
its main features responsible for the bending correspond to reality.
At the same time, the real situation is certainly more complex.

\subsection*{Results Poorly Agree with Earlier Theories of Bending}

All theories proposed during the last 20 years to explain intrinsic
bends in DNA double helices agree with some experimental observations
and disagree with the other and, probably, each of them continues to
attract some proponents. Here we compare our results with the most
popular models of bending regardless of their experimental validation.
Comparisons with experimental data have been the subject of many
reviews
\cite{Crothers:99,Diekmann:87a,Hagerman:90,Crothers:90,Crothers:92,Olson:96}.

The wedge model of DNA bending \cite{Trifonov:80} resulted from
merging of ideas developed in seventies to explain the ability of a
double helix to wrap around nucleosome particles.  The first idea was
that this can occur due to kinks of the helical axis phased with the
helical screw \cite{Crick:75,Zhurkin:79}, with kinks implying
destacking of base pairs in fable points in order to maintain perfect
stacking elsewhere. The second idea was that the double helix can be
smoothly bent, without destacking, by small deformations in every base
pair step \cite{Namoradze:77}. The wedge model merges the two by
postulating that, in every specific dinucleotide, the preferred
stacking of bases is slightly non-parallel and this causes bending in
the same way as kinks do. It can be further developed by increasing
the number of wedge degrees of freedom, by considering triplets,
tetraplets, and so forth instead of dinucleotides, and by assuming
that the non-zero average wedges result from random sampling from
asymmetrical energy valleys around local energy minima, rather than
from minimum energy configurations \cite{Maroun:88}. Depending upon
the specific wedge parameters, this model can place the curvature
inside A-tracts or between them \cite{Maroun:88,Goodsell:94} and also
explain bending in non A-tract sequences \cite{Bolshoy:91}. For the
present discussion, it is convenient to unite all such mechanisms in
one group characterized by the tacit emphasis upon the specific base
pair stacking preferences as the source of the DNA bending.

The results shown in Figs. \ref{Fhlpa} and \ref{Fhlpawn} obviously
disagree with these views. There is little similarity between matching
dinucleotides in the same structure and, moreover, base pair steps put
in the same sequence context in three closely similar bent
conformations exhibit broadly different helical parameters. The last
observation means that even a generalized wedge model with
dinucleotide blocks replaced by triplets, tetraplets, and so forth,
would disagree with our results.

The junction model \cite{Levene:83} postulates that there is a
distinct specific A-tract form of the double helix
characterized by a stronger inclination of base pairs with respect to
the helical axis than in the normal B-DNA. In this case, planar
stacking at the junction between the two structures would result in a
kink of the helical axis. Formally, such geometry can also be obtained
with the generalized wedge model above, but the junction theory puts
an emphasis upon the specific A-tract form of DNA as the principal
physical cause of bending.  Its structure can be due to cooperative
interactions in long DNA stretches and its environment.  Within the
framework of the junction model, particular roles were sometimes
attributed to bifurcated hydrogen bonds \cite{Aymami:89}, the water
spine in the minor groove \cite{Chuprina:88}, or the NH$_2$ groups of
adenines \cite{Kitzing:87}.

It is evident that the junction model also poorly agrees with our
results. In dynamics, conformations, both smoothly bent and kinked at
the two insertions between the A-tracts, are observed periodically.
The kinks, however, are not centered at the boundaries between
A-tracts and the flanking sequences, and they are not sharp.  In such
conformations, A-tracts are less bent than regions between them, that
is the bend is localized but still smooth. In Fig. \ref{Fhlpawn} the
inclination shows smooth oscillations, even without window averaging,
with no kinks. It decreases from 5' to 3' ends of A-tracts, and since
the 3' ends of A-tracts are dephased and positioned differently with
respect to the bending plane, no evident relationship to bending can
be readily seen. All helical parameters vary along the sequence so
that there is no A-tract fragment where they repeat at two consecutive
steps. Thus, although the structures are bent, the specific regular
A-tract structure is not seen, as well as the ``random B-DNA'',
though, which are the two key components of the junction model.

The third model, which was first mentioned in the context of the
junction theory \cite{Levene:86}, but became popular only in the
recent years \cite{Hud:99}, attributes the cause of bending to solvent
counterions. If they are specifically bound by minor grooves of
A-tracts, in a phased sequence, phosphate groups would appear
partially neutralized at one side of the double helix, and the
repulsion between the opposite phosphates would bend DNA towards minor
grooves of A-tracts. The very fact that the minimal model of B-DNA,
without explicit counterions, produces static bends, in good agreement
with experiments, strongly suggests that the counterions hardly play a
key role in the A-tract induced bending.

At the same time, our results do not contradict less specific
non-local theories of A-tract bending. The modified junction model
\cite{Nadeau:89} assumes that the deformations at the boundary between
the two conformations can propagate for several base pair steps. The
A-tracts in the sequence studied here may be too short for their
ingenious structure to establish. The second such theory
\cite{Burkhoff:87} proposed that the bending is caused by the
modulations of the minor groove. Really, the double helix is usually
bent towards the major groove at the minor groove widenings, and in
the opposite direction at its narrowings. In TJA, for instance, this
relationship is maintained during the rotation of the bending plane.
Sometimes, however, the double helix straightens and remains unbent
during nanoseconds, while the minor groove profile does not change
\cite{Mzlanl:00}.

It is understood, however, that the last two non-local models are
incomplete.  Actually, they cannot be verified or disproved because
the issue of the physical origin of bending is tacitly dropped. Simple
geometrical considerations dictate that the grooves must be narrower
at the inside edge of the bend \cite{Drew:85}. One may postulate,
therefore, that groove modulations cause bending or, {\em vice versa},
that it is bending that causes groove modulations, but the physical
origin of the phenomenon remains obscure. Similarly, the modified
junction model essentially discards the essence of the original
theory, which considers the specific poly-dA structure as the source
of the bend.  If the ``boundary deformations'' can exist without the
structures and boundaries themselves one should look for another force
that maintains these deformations.

Generally speaking, the results presented here are best interpreted if
one assumes that there is an external force that imposes a bent
shape upon the double helix as a series of mechanical constraints. The
double helix is allowed to move, but so that these constraints would
remain fulfilled. Thus, the bases can change their mutual orientation
and the backbone can switch between $\rm B_I\ and\ B_{II}$
conformations, but the overall proportion of the $\rm B_{II}$
conformers remains constant, and fluctuations of helical parameters in
the neighboring base pair steps tend to compensate.

\subsection*{Possible Physical Origin of Spontaneous Static Bends in
Double Helices}

The hypothesis outlined below is based upon our computational
results as well as upon analysis of well-known experimental data.
Although it does not answer all unclear questions concerning DNA bending
we consider it most likely and describe it here for discussion and
further investigation.

Let us first ask the following simple geometric question: ``What is
the shortest line that joins two points on a surface of a cylinder?''
To answer it, one should first cut the cylinder parallel to its axis,
unfold its surface onto a plane, join the two points by a straight
line and then fold the surface back upon the cylinder. The resultant
curve represents an interval of a spiral trace with a constant
inclination to the cylinder axis. Now consider an ideal canonical
B-DNA model of a double helix. The stacked base pairs form the core of
a cylinder and the sugar-phosphate backbone forms an ideal
spiral trace on its surface, that is the shortest line that
joins the ``surface'' nitrogens $\rm N_1/N_9$ of the bases. If we now
assume that the backbone is a stiff elastic that can be characterized
by a certain specific length, we are obliged to conclude that this
model implicitly assumes that the backbone is stretched and tends to
reduce its length on the surface of this cylinder. Our last question
is: ``What would happen if the preferred backbone length appeared to be
longer that allowed by the canonical model?'' A simple answer is: it would
try to extend by pushing bases. They can accommodate this extention
within the framework of a regular helical structure by increasing the
helical twist and changing other helical parameters. This option,
however, is opposed by the loss in the stacking energy and, when it
becomes difficult to extend in this way, the backbone will try to
deviate from the ideal spiral trace. In this case the
the grooves on the surface of the Watson-Crick double helix can no
longer maintain a constant width.

It seems possible to assume that, in physiological conditions, the
equilibrium specific length of the DNA backbone is actually larger than
that allowed by a regular B-DNA structure with the average helical
twist of 34.5\degree. Its further extention is opposed by the limit of
the tolerance of the pase pair stacking and, as a result, the backbone
appears ``frustrated'' and is forced to wander on the cylindrical
surface formed by base pair stack. The ideal parallel stacking has to
be perturbed and we believe that it is this effect that eventually
bends the double helix. Modulations of the DNA grooves, which is a
well-known ubiquitous feature of the single crystal B-DNA structures,
is a natural indicator of this particular state of the backbone. It
is observed for very different sequences as an apparently general
attribute of the B-DNA form in physiological conditions. Thus, if we
had to decide whether the DNA backbone in stretched, relaxed or
compressed by looking only at the single crystal B-DNA structures, we
would be obliged to conclude that the first option looks unlikely, the
second is possible, while the third is the most probable. A compressed
backbone is more likely to cause smooth groove modulations found in
experimental structures than a relaxed one, which would rather be
controlled by the local sequence effects.

Let us consider an ideal B-DNA model, with planar base pairs
perpendicular to the helical axis, and try to imagine how wandering of
the backbone traces can emerge. For simplicity, we first consider the
helical twist as the only variable parameter. Obviously, by smoothly
increasing and decreasing the twist we obtain, respectively,
narrowings and widenings of the minor groove. The desired backbone
waving emerges, and a larger its length can be accommodated on the same
cylindrical surface. It is easy to see, however, that, if the
parallel stacking is maintained, the backbone must be compressed when
the twist is reduced and stretched in the opposite phase. In reality,
however, the backbone is stiff and it cannot be compressed
significantly, therefore, it is the stacking that suffers when the
twist is reduced. Although this description is simplistic,
and other base pair degrees of freedom in addition to the twist
can contribute to the wandering, it captures
the essence of the underlying mechanics. In the widenings of the minor
groove, where the twist is reduced, the backbone pushes off the
neighboring base pairs at C1' atoms, causing various perturbations of
the parallel stacking. On average, they are likely to deviate the
helical axis towards the major groove because C1' atoms are at the
minor groove side. These perturbations are delocalized and involve
rolling, tilting, as well as other relative motions of base pairs,
and there is an ensemble of orientations that fulfill the
constraints imposed by the backbone lengths, rather than a single
preferred bent conformation. The modulations of the minor groove
width and the bending of the double helix appear related, as
was suggested earlier \cite{Burkhoff:87}, because they represent two
consequences of a single cause. They are related as brothers rather
than as a parent and a child and, probably, are not bound
to always appear together.

The major component of the backbone stiffness is the electrostatic
repulsion between the charged phosphate groups. Even though this
repulsion is shielded by water and counterions, the experiments where
bending in B-DNA was induced by specific neutralization of phosphates
\cite{Strauss:94} proved that they are not shielded even when separated
by two helical turns. Complete neutralization, therefore, is hardly imaginable.
The local electric field around a pair of neighboring phosphates in
the same strand is created by all surrounding charges, including
the phosphates of the opposite strand, and it favors maximal
possible separation between $\rm P_n\ and\ P_{n+1}$.
In B-DNA, this distance is close to the absolute maximum, which
is achieved by putting all relevant backbone torsions except
one in the {\em trans} configuration \cite{Saenger:84}.
The maximal extention gives the ground energy state with the $\rm P_n -
 P_{n+1}$ distance around 7.7 \AA. The corresponding thermodynamic
average for a free backbone in solution is $D(T,\epsilon)<7.7$ \AA,
where $T$ is the temperature, and $\epsilon$ is the effective
dielectric constant that depends upon the concentration of
counterions. In the single crystal structures, the largest $\rm P_n -
P_{n+1}$ distances observed are in the range of 7.3 -- 7.6 \AA\
suggesting that there are no prohibitive steric obstacles for a
completely extended backbone. At the same time, the distances most
commonly found are around 6.7 \AA\ while in the fiber canonical
structure it is 6.5 \AA. Apparently, with normal temperature in
physiological conditions $D(T,\epsilon)\gtrsim 6.7$, and the backbone
is forced to wander. $D(T,\epsilon)$ should be a decreasing
function of both arguments, therefore, the backbone stiffness and,
accordingly, the curvature should decrease as the temperature grows
and/or as the phosphate shielding is improved by increasing the ionic
force. These two non sequence-specific effects have been found in
experiments \cite{Diekmann:85}.  With $D(T,\epsilon)\approx 6.5$ the
backbone relaxes and the phenomenon of DNA bending vanishes.

According to this hypothesis, the transition of A-tracts in a specific
DNA form is not an indispensable prerequisite of the bending. Moreover,
it suggests that the regular poly-dA structure may not exist in
solution because, with the average twist increased to 36\degree, the
backbone is, possibly, still compressed and continues to wander,
although with a longer characteristic wave lengths.  The structures of
short A-tracts computed here are rather variable and it is not clear
how they can be extended to longer poly-dA double helices. We believe
that the A-tracts rather label the regions where higher twist values
are allowed by the base stacking interactions. The backbone prefers
to compress the minor groove here, thus fixing the phase of its
modulations along the double helix. In random and homopolymer
sequences the minor groove probably also narrows and widens, but the
corresponding maxima and minima can migrate along the double helix in
a way similar to that observed here during the rotation of the bending
plane in TJA.

Our model considers the bending of a DNA double
helix as a deformation imposed upon the stem of the stacked base pairs
by interactions external to it. The bases are forced to ``forget''
their preferred stacking orientations and look for a possibility to
maintain the overall structure by sampling the orientations at the
limit of destacking. At the same time, it is the broad ``tolerance''
of the base pair stacking that makes all this game possible.
If true, this theory gives a slightly different overall
view of the DNA molecule in physiological conditions and
entails important biological consequences. Notably, it
suggests that the local DNA structure is not simply determined by the
stacking preferences of base pairs in dinucleotides, trinucleotide,
and so forth. The two waving backbone profiles on the surface of
the helix impose a medium range context upon the local base pair
stacking because the phases of these modulations can well
correlate over several DNA turns. This makes possible mutual
dependence of local conformations in sites separated by
considerable DNA stretches. Fine tuning of phases of these modulations
may be the function of single short A-tracts as well as of some regulatory
proteins. The degree of the backbone compression is connected with
supercoiling and can be controlled in this way, which gives yet
another possible instrument of structural regulation.
One may note also that this theory offers a unified model which
explains static bends in A-tract and non A-tract sequences
as well as the bending induced by the negative supercoiling
in circular DNA.

\subsection*{Conclusions}

The static curvature spontaneously emerges in free MD simulations of an
atom level molecular model of B-DNA double helix, with the nucleotide
sequence as a single structural bias. Convergence to similar
statically bent states have been demonstrated in three independent MD
trajectories of 10-20 ns. The bending direction and its magnitude are
in good agreement with experimental observations.  Unexpectedly, the
three computed MD structures exhibit a striking microscopic
heterogeneity as regards variations of helical and conformational
parameters along the molecule. Based upon the computational results as
well as the literature experimental data a new possible mechanism of
bending in a double helical DNA is proposed. It postulates that, in
physiological conditions, the equilibrium specific length of the
DNA backbone is larger than is admissible in the regular B-DNA form, which
forces it to fold in a wavy trace on the cylindrical surface of the
double helix. This results in modulations of the minor groove
width, slight asymmetrical destacking of base pairs at the groove
widenings and, eventually, in bending of the DNA molecule.

\subsection*{Methods}

Molecular dynamics simulations have been performed with the internal
coordinate method (ICMD) \cite{Mzjbsd:89,Mzjcc:97} including special
technique for flexible sugar rings \cite{Mzjchp:99}. The so-called
minimal B-DNA model was used \cite{Mzjacs:98,Mzlanl:99} which consists
of a double helix with the minor groove filled with explicit water. It
does not involve explicit counterions and damps long range
electrostatic interactions in a semi-empirical way by using linear
distance scaling of the electrostatic constant and reduction of
phosphate charges. The DNA model was same as in earlier reports,
\cite{Mzjacs:98,Mzlanl:99} namely, all torsions were free as well as
bond angles centered at sugar atoms, while other bonds and angles were
fixed, and the bases held rigid.  Molecular dynamics calculations were
carried out with a time step of 10 fsec and the effective inertia of
planar sugar angles increased by 140 amu$\cdot$\AA$^2$ as explained
elsewhere \cite{Mzjacs:98}.  The coordinates were saved once in 2.5
ps. AMBER94 \cite{AMBER94:,Cheatham:99} force field and atom
parameters were used with TIP3P water \cite{TIP3P:} and no cut off
schemes.

The initial conformation for TJBa was prepared by vacuum energy
minimization starting from the fiber B-DNA model constructed from the
published atom coordinates \cite{Arnott:72}. The subsequent hydration
protocol to fill up the minor groove \cite{Mzjacs:98} normally adds
around 16 water molecules per base pair. The starting state for TJBb
was obtained by energy minimizing an equilibrated straight structure
taken from the initial phase of TJBa. The initial conformation for TJA
was prepared by hydrating the minor groove of the corresponding A-DNA
model \cite{Arnott:72} without the preliminary energy minimization. In
TJA, the necessary number of water molecules was added after
equilibration to make it equal to that in TJBa and TJBb.

The heating and equilibration protocols were same as before
\cite{Mzjacs:98,Mzlanl:99}. During the runs, after every 200 ps, water
positions were checked in order to identify those penetrating into the
major groove and those completely separated.  These molecules, if
found, were removed and next re-introduced in simulations by putting
them with zero velocities at random positions around the hydrated
duplex, so that they could readily re-join the core system. This
procedure assures stable conditions, notably, a constant number of
molecules in the minor groove hydration cloud and the absence of water
in the major groove, which is necessary for fast sampling
\cite{Mzlanl:99}. The interval of 200 ps between the checks is small
enough to assure that on average less then one molecule is
repositioned and, therefore, the perturbation introduced is considered
negligible.

\section*{Appendix}
This section contains comments from anonymous referees of a peer-review
journal where this and a closely related paper entitled {\em ``Molecular
Dynamics Studies of Sequence-directed Curvature in Bending Locus
of Trypanosome Kinetoplast DNA''} has been considered for
publication, but rejected.
\subsection{Journal of Molecular Biology}

\subsubsection {First referee}

These companion manuscripts describe a series of molecular dynamics
trajectories obtained for DNA sequences containing arrangements of oligo
dA - oligo dT motifs implicated in intrinsic DNA bending.  Unlike previous
MD studies of intrinsically bent DNA sequences, these calculations omit
explicit consideration of the role of counterions.  Because recent
crystallographic studies of A-tract-like DNA sequences have attributed
intrinsic bending to the localization of counterions in the minor groove, a
detailed understanding of the underlying basis of A-tract-dependent bending
and its relationship to DNA-counterion interactions would be an important
contribution.

Although the MD calculations seem to have been carried out with close
attention to detail, both manuscripts suffer from some troubling problems,
specifically:

The DNA sequence in question is a 25-bp deoxyoligonucleotide that contains 3
A/T tracts.  Two of these are arranged in phase with the helix screw with
the third tract inverted with respect to the other two.  Extrapolating from
available experimental data, this sequence is expected to confer some degree
of intrinsic bending.  The main focus of this manuscript is the comparison
of data obtained for an MD trajectory computed from an A-form starting
conformation (TJA) with two other trajectories that begin with B-form
structures (TJBa and TJBb).

Significant differences in behavior and in time-averaged helical parameters
are observed for the TJA trajectory compared with both TJBa and TJBb,
suggesting that the structures are not fully equilibrated.  This is
particularly evident in the computed bending direction, which varies
dramatically during early times in the TJA trajectory.  Even after 15 ns,
when the orientations of bending planes appear to have approached asymptotic
values, the TJA plane is displaced by between 30 and 60 degrees from those
of TJBa and TJBb, which are quite similar to one another.  This fact
strongly suggests that the MD-simulation results depend nontrivially on
initial conditions, even after 15-20 ns, which calls into question most of
the results obtained from the computed trajectories.

\subsubsection{Second referee}

        Dr. Mazur reports the results of MD simulations of DNA 25-mers
sequences that contain three phased A-tracts. He believes that he has
obtained the first model system in which properly directed static curvature
emerges spontaneously in conditions excluding any initial bias except for
base pair sequence. He observes that the ensemble of curved conformations
reveals significant microscopic heterogeneity, which he believes is in
contradiction to existing theoretical models of DNA bending. In CAM110/00
he performs a series of simulations on a DNA fragment that has not been
shown experimentally to bend in solution. In this case the DNA sequence was
chosen based on its propensity to adopt a characteristic structure during
simulations. In CAM167/00 he performs a similar investigation on B-DNA
fragment composed of a sequence that has been shown experimentally to bend.
My view is these two papers should be combined as one, and the review will
treat them as one.

        I found this paper to be interesting and possibly worthy of
publication in JMB, even as I took issue with a substantial portion of it.
The basic premise of the paper is that a model lacking realistic
electrostatics can provide meaningful information about long range DNA
conformation. In Mazur's model, long range electrostatic interactions are
dampened and phosphate charges are attenuated.

        I had some concerns about the basic rationale for the
non-electrostatic model.  I initially assumed that it's greater simplicity
would allow longer trajectories, etc. But the trajectories of Dr. Mazur are
not substantially longer than those described by Beveridge, Pettitte, etc.
And in fact that seems not to be the rationale. Dr. Mazur believes that
full atom force fields, with explicit ions, give less realistic results
than his electrostatic-attenuated model. In particular he says that full
atom force fields give slightly overwound DNA, which camouflages  DNA
bending. What is the cause of this? A problem in the force field? Why not
fix that instead of going the non-electrostatic route? I am not comfortable
enough with the world of MD pass judgment on this issue, but think someone
who is should evaluate that prior to publication.

        I just went back and re-read Diekmann's classic 1985 JMB paper
[Diekmann, S., \& Wang, J. C. "On the sequence determinants and flexibility
of the kinetoplast DNA fragment with abnormal gel electrophoretic
mobilities" (1985) J. Mol. Biol. 186, 1-11.] Diekmann shows clearly that
the electrophoretic anomaly of kinetoplast DNA decreases with increasing
Na, and increases very dramatically with increasing Mg. His experiments
seem well-conceived , well-conducted and well-analyzed. For example he
implants a temperature sensor within his gels to insure constant, fixed
temperature. One has to believe Diekmann's results. My fundamental problem
with Mazur's model is that it cannot account for experimental data. How can
bending be cation-dependent, but the mechanism not be electrostatic in
nature?

        Mazur does concede that experimentally "curvature is reduced in
high salt, but for some sequences it is increased in the presence of
divalent metal ions" (cites Diekmann). [page 2 MS CODE CAM110/00]. But the
implication here is that the observation of Diekmann is not general to all
A-tracts. The next sentence of the manuscript may be read as confirming
that the cation effect is not general, but is length and composition
dependent (the text is a little confusing here). However the Woo and
Crothers citation, used as support for that, does not discuss the cation
effect.  If there are data somewhere suggesting that the cation effects are
not general, they should be cited and discussed. That would really increase
the strength of Mazur's argument. If not, the text should be clarified.

        An additional issue that is not illuminated  much here is the
comparison of Mazur's model with the results of x-ray crystallography. In
crystals of oligonucleotides, A-tracts are straight ("less prone to bending
than other sequences" is rather understating it). To accommodate this
observation, Dickerson (JMB 1994) proposed a model in which A-tract DNA
curvature results from roll-bending of non-A-tracts, and linear A-tracts.
Crothers (JMB 1994) is contemptuous of that model, and believes that the
linear conformations of A-tracts observed thus far in crystals are not
those associated with the curvature 'observed' in gel mobility experiments.
In fact such a discrepancy between dilute solution (where intramolecular
forces would dominate) and condensed states (such as crystals, where
intermolecular forces dominate) is expected if long-range electrostatics
play a key role in curvature. Those long range forces are turned off in
Mazur's model. (He does seem to allude to the crystallography/solution
discrepancy on page 19). So again his model does not account for
experimental data.

        As an aside: Mazur believes that groove narrowing and bending are
coupled. How does one then explain the observation that A-tracts in
crystals have narrow minor grooves, yet are not bent?

        Finally, some aspects of Mazur's (combined) model seems to be
inconsistent and self-contradictory. In his model (as I understand it), (1)
electrostatic repulsion between adjacent phosphate groups drives helical
twisting, (2) A-tracts are regions where higher helical twist is
facilitated by lower stacking energies in comparison to those of G-C base
pairs, (3) higher helical twist narrows the minor groove, and (4) groove
narrowing is somehow related to axial curvature (this is a little unclear;
the description "as brothers rather than a parent and a child" did not
enlighten me).  This model has certain attractive features, [the idea that
electrostatic repulsion between adjacent phosphate groups drives helical
twisting while stacking opposes it was previously presented by Alex Rich in
1992 in a chapter of Structure \& Function, Volume I: Nucleic Acids pp.
107-125 (from a Sarma meeting)] but some deficiencies also. If
electrostatic repulsion between adjacent phosphate groups drives helical
twisting, then how can correct values of helical twist be obtained with
attenuated phosphate charges? Or restated: Does this model not ascribe
electrostatic forces as the ultimate cause of static bends, contradicting
the non-electrostatic assumption? And I just checked in one crystal
structure and found a place where OP to OP (phosphate oxygens, where the
negative charge resides) across the minor groove are less than those
between adjacent phosphates. How can electrostatic repulsion between
adjacent phosphate groups drive other phosphate groups together like that,
especially if stacking forces are working in opposition? How can one
understand such phenomena without explicitly considering electrostatic
interactions?

        Although the bulk of this review might appear rather critical, a
model can be useful even if it does not account for all data. And that may
be the case here.  If Mazur has indeed obtained the first model system
where properly directed static curvature emerges spontaneously, then his
model clearly has utility. If a reviewer who specializes in MD simulations
(not this reviewer) would confirm that, and support the utility of the
approach, then publication may be in order. However I would like the paper
more if it were reformulated as an exploration of possible models rather
than the last word on the physical origin of intrinsic bends.

Re: measurement of the groove width: Is the some reason that an old version
of Curves was used? The newer versions fit a surface to the groove, rather
than just measure phosphate-phosphate distances, and provide a much finer
view of groove width.

\end{document}